\documentclass[10pt,letterpaper,twocolumn]{article} 

\usepackage{ol2}
\usepackage[draft]{hyperref}
\usepackage{amsmath}
\usepackage{euscript}
\usepackage{graphics}

\begin{document}

\twocolumn[

\title{Goos-H\"{a}nchen and Imbert-Fedorov shifts of polarized vortex beams}

\author{Konstantin Y. Bliokh,$^{1,2,*}$ Ilya V. Shadrivov,$^1$ and Yuri S. Kivshar$^1$}

\address{$^1$Nonlinear Physics Center, Research School of Physical Sciences and
Engineering, \\ Australian National University, Canberra ACT 0200, Australia
\\ $^2$Institute of Radio Astronomy, 4 Krasnoznamyonnaya Str., Kharkov 61002, Ukraine\\
$^*$Corresponding author: k.bliokh@gmail.com}

\begin{abstract}
We study, analytically and numerically, reflection and transmission of an arbitrarily
polarized vortex beam on an interface separating two dielectric media and derive general
expressions for linear and angular Goos-H\"{a}nchen and Imbert-Fedorov shifts. We predict
a novel vortex-induced Goos-H\"{a}nchen shift, and also reveal direct connection between
the spin-induced angular shifts and the vortex-induced linear shifts.
\end{abstract}


]

\noindent Reflection and transmission of plain waves at an interface separating two
homogenous isotropic media are described by the well-known Snell's and Fresnel equations.
However, the problem becomes tricky for confined beams with a finite spectral
distribution. Interference of partial plain waves propagating at slightly different
angles and obeying individual Snell and Fresnel formulas, results in such effects as the
longitudinal Goos-H\"{a}nchen (GH) \cite{GH,Artmann} and transverse Imbert-Fedorov
(IF)~\cite{Fedorov,Imbert} shifts, which displace the output beams within and across the
propagation plane, respectively (see Fig. 1).

While the GH shift was explained and calculated soon after its discovery~\cite{Artmann},
the transverse IF shift was associated by significant controversies over about 50 years.
The direct calculation of the IF effect was first performed for the reflected
beam~\cite{Schilling,Ricard, Hugonin} and later generalized to the transmitted
beam~\cite{Fed1985, Zel}. It was shown that the IF shift is closely related to the spin
angular momentum carried by a polarized beam and conservation of the total angular
momentum in the system~\cite{Player, Fed1988} (see also~\cite{OMN,Bliokh1,Bliokh2}).
Despite long history of the theoretical studies and
experiments~\cite{Imbert,Floch,Gilles}, analytical expression for the IF shift of a
polarized Gaussian beam was derived in the correct form only recently~\cite{Bliokh1,
Bliokh2}, and these results have been confirmed both experimentally~\cite{Hosten} and
theoretically~\cite{AW}.

In addition to the usual linear shifts, \emph{angular} GH and IF shifts caused by the
beam diffraction have been described~\cite{Bliokh2, AW}. Recently, this description of
the IF effect has been extended to the case of higher-order \textit{vortex beams}
carrying intrinsic orbital angular momentum~\cite{Fed2001}. The vortex-induced IF shift
is proportional to the vortex charge, but it also significantly depends on the beam
polarization. Such IF shift was calculated~\cite{Fed2001} and measured
experimentally~\cite{DG} for $p$ and $s$ polarizations only, when the spin IF effect
vanishes.

The main purpose of this Letter is twofold. First, we derive explicit analytical
expressions for both linear and angular IF and GH shifts in the most general case of {\em
an arbitrarily polarized vortex beam}. We unveil a direct relation between the
vortex-dependent IF shift and the angular GH shift~\cite{AW} and predict novel
\textit{vortex-induced GH shift} related to the angular IF shift. Second, we verify all
theoretical results by direct numerical simulations.

We consider the reflection and refraction of an optical beam at an interface separating
two media, as shown in Fig. 1. In addition to the coordinate system $(x,y,z)$ attached to
the interface, we employ the coordinate systems of individual beams $(X^a,Y^a,Z^a)$,
where $a=i,r,t$ denotes incident, reflected and transmitted beams, respectively. The
$Z^a$ axis attached to the directions of the $a$-th beam as determined by the Snell's
law. The incident beam propagates in the $(x,z)$ plane, so that $Y^a=y$ (see Fig.~1). We
also define the wave number in the first medium, $k$, the angle of incidence, $\theta$,
the angle of refraction, $\theta^{\prime}=\sin^{-1}(n^{-1}\sin\theta)$, as well as the
relative permittivity $\varepsilon$, permeability $\mu$, and refractive index
$n=\sqrt{\varepsilon\mu}$ of the second medium.

\begin{figure}[t]
\centering \scalebox{0.33}{\includegraphics{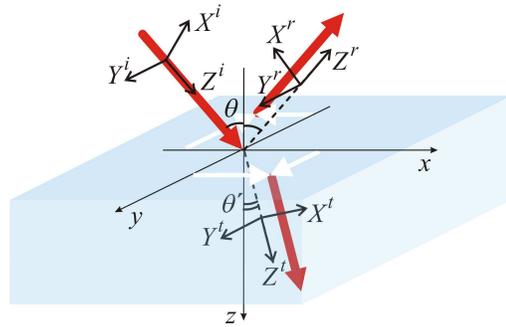}}
\caption{(Color online) Geometry of the beam reflection and transmission at an interface. Projections of the GH shifts, $- \langle X^{r} \rangle \cos\theta$, $\langle X^{t} \rangle \cos\theta^{\prime}$, and the IF shifts, $\langle Y^{r,t} \rangle $, are shown by white arrows.}
\label{Fig1}
\end{figure}

We assume that the incident beam is a uniformly polarized Laguerre-Gaussian beam with the
waist located at the interface, so that the transverse $(X^i,Y^i)$-component of its
electric field has the form
\begin{equation}
{\bf E}_{\rm trans}^i\propto \left(e_{\parallel} {\bf \hat X}^i+e_{\perp} {\bf \hat Y}^i\right){\EuScript L}_{n,m}(X^i,Y^i,Z^i).
\end{equation}
Here ${\EuScript L}_{n,m}$ is the Laguerre-Gaussian solution of the scalar parabolic wave
equation which contains an optical vortex of charge $m$: ${\EuScript L}_{n,m}\propto
\left(X^i+i\,\mathrm{sign}\, (m)\, Y^i\right)^{|m|}$ \cite{OAM}, whereas polarization
components form the normalized Jones vector in the basis of $p$ and $s$ modes:
$(e_{\parallel},e_{\perp})^T$, $|e_{\parallel}|^2+|e_{\perp}|^2=1$.

The main characteristics of the reflected and transmitted beams are determined by the
Fresnel coefficients, $R_{\parallel}$, $R_{\perp}$, $T_{\parallel}$, and $T_{\perp}$. The
amplitude reflection and transmission coefficients $R,T$ and the corresponding energy
reflection and transmission coefficients $Q^{r,t}$ are given by
\begin{eqnarray}
\nonumber
R=\sqrt{|R_{\parallel}e_{\parallel}|^2+|R_{\perp}e_{\perp}|^2},~T=\sqrt{|T_{\parallel}e_{\parallel}|^2+|T_{\perp}e_{\perp}|^2},\\
Q^r=|R|^2,~Q^t= \frac{n\cos\theta^{\prime}}{\mu\cos\theta}|T|^2,~Q^r+Q^t=1,
\end{eqnarray}
The Jones vectors of the secondary beams are
\begin{equation}
\left(\begin{array}{c}e_{\parallel}^r\\e_{\perp}^r\end{array}\right)=\frac{1}{R}\left(\begin{array}{c}R_{\parallel}e_{\parallel}\\R_{\perp} e_{\perp}\end{array}\right),~
\left(\begin{array}{c}e_{\parallel}^t\\e_{\perp}^t\end{array}\right)=\frac{1}{T}\left(\begin{array}{c}T_{\parallel}e_{\parallel}\\T_{\perp} e_{\perp}\end{array}\right).
\end{equation}

To derive the lateral shifts of an arbitrarily polarized vortex beam, we use the basic
results obtained for polarized Gaussian beams ($m=0$). In the regime of the total
internal reflection, $\sin\theta > n$, the Fresnel reflection coefficient become complex:
$R_{\parallel,\perp}=\exp(i\varphi_{\parallel,\perp})$, and the GH, $\langle X \rangle$,
and IF, $\langle Y \rangle$, shifts are described by the Artmann and Schillings formulas
\cite{Artmann,Schilling}:
\begin{eqnarray}
\langle X^r \rangle _{tot}= \frac{1}{k}\left(|e_{\parallel}|^2\frac{\partial \varphi_{\parallel}}{\partial\theta}+|e_{\perp}|^2\frac{\partial \varphi_{\perp}}{\partial\theta}\right),\\
\langle Y^r \rangle _{tot}= -\frac{\cot\theta}{k} \left[\sigma (1+\cos\delta)+\chi\sin\delta\right],
\end{eqnarray}
where $\sigma=2{\rm Im}(e_{\parallel}^* e_{\perp})$ is the helicity of the incident beam
(degree of circular polarization), $\chi=2{\rm Re}(e_{\parallel}^* e_{\perp})$ is the
degree of linear polarization inclined at $\pi/4$ with respect to the incident plane, and
$\delta=\varphi_{\perp}-\varphi_{\parallel}$. We note, that the shifts here and below in
the paper are given in the coordinate system of the respective beam.

In the regime of partial reflection and transmission, $\sin\theta < n$, the Fresnel
coefficients are real, and the GH shift vanishes, $\langle X^{r,t} \rangle=0$, while the
IF shifts are given by the equations derived in \cite{Bliokh1,Bliokh2}:
\begin{eqnarray}
\nonumber
\langle Y^r \rangle _0 = -\sigma \frac{\cot\theta}{2k}\frac{(R_{\parallel}+R_{\perp})^2}{R^2},\\
\langle Y^t \rangle _0 = -\sigma \frac{\cot\theta}{2k}\frac{T_{\parallel}^2+T_{\perp}^2-2T_{\parallel}T_{\perp}\cos\theta^{\prime} /\cos\theta}{T^2}.
\end{eqnarray}
In addition, the reflected and transmitted beams undergo \textit{angular shifts}, which
can be considered as the shifts in the \textit{wave vector} space. Generalizing results
of \cite{Bliokh2,AW}, these shifts can be written as
\begin{eqnarray}
\langle K_{x}^{r} \rangle_0 = - \frac{1}{2D}\frac{d \ln Q^r}{d\theta},~
\langle K_{x}^{t} \rangle_0 = \frac{\cos\theta}{2D\cos\theta^{\prime}}\frac{d \ln Q^t}{d\theta},\\
\langle K_{y}^{r} \rangle_0 = \chi \frac{\cot \theta}{2D}\frac{R_{\parallel}^2-R_{\perp}^2}{R^2},~
\langle K_{y}^{t} \rangle_0 = \chi \frac{\cot \theta}{2D}\frac{T_{\parallel}^2-T_{\perp}^2}{T^2}.
\end{eqnarray}
Here we denote $K^a_{x}$ and $K^a_{y}$ as the $X^a$ and $Y^a$ wave vector components in
the $a$-th beam, $D=kw_0^2/2$ is the Rayleigh length, $w_0$ is the minimum beam waist,
and the derivatives of the energy reflection and transmission coefficients (7) are
explicitly calculated (for the $p$ and $s$ polarizations) in \cite{Fed2001}.

Now we consider a polarized incident beam with vortex, $m\neq 0$. Such beam carries \textit{intrinsic orbital angular momentum} (AM) ${\bf L}^i=m{\bf \hat Z}^i$  (in units of $\hbar$) per one photon~\cite{OAM}. From the Snell's laws it follows that the AMs of the reflected and refracted beams are given by \cite{Fed2001,Fed1}
\begin{equation}
{\bf L}^r=-m{\bf \hat Z}^r,~{\bf L}^t=\frac{1}{2}\left(\frac{\cos\theta}{\cos\theta^{\prime}}+\frac{\cos\theta^{\prime}}{\cos\theta}\right)m{\bf \hat Z}^t.
\end{equation}
As was shown by Fedoseyev~\cite{Fed1} for a particular cases of $p$ and $s$ polarization,
the IF shift of vortex beams consists of two contributions. The first one is \textit{a
polarization-independent} shift that comes from the difference in the $z$-components of
intrinsic AM at the reflection and refraction which is compensated by the transverse
shift producing an \textit{extrinsic} AM \cite{Player,Fed1988,OMN,Bliokh1,Bliokh2}:
$\langle Y^{r,t} \rangle _{1} =(L^{r,t}_z-L^i_z)/k\sin\theta$. Taking into account that
$L^{i}_z=L^{i}\cos\theta$, $L^{r}_z=-L^{r}\cos\theta$, and
$L^{t}_z=L^{t}\cos\theta^{\prime}$, one obtains \cite{Fed2001,Fed1}:
\begin{equation}
\langle Y^{r} \rangle _{1}=0,~ \langle Y^{t} \rangle _{1}=\frac{m}{2k}\tan\theta (1-n^{-2}).
\end{equation}

The second contribution is essentially \textit{polarization-dependent}. Here we
demonstrate that it is directly related to the angular GH shift (7). Indeed, as is shown
in \cite{AW}, the angular shift (7) induces \textit{an imaginary} shift in the Gaussian
envelope of the beam: $X^{r,t}\rightarrow X^{r,t}-iD^{r,t}_x\langle K_{x}^{r,t} \rangle_0
/k$ (where $D^r_x=D$ and $D^t_x=D\cos^2\theta^{\prime}/\cos^2\theta$ \cite{Bliokh2}). It
can be readily seen that this imaginary shift produces a \textit{real} shift of the
vortex in the \textit{orthogonal} direction:
\[
\left[\gamma^{r,t} X^{r,t}+i\,{\rm sign}\, (m) \left(Y^{r,t}-\gamma^{r,t}\frac{D^{r,t}_x}{k}\langle K_{x}^{r,t} \rangle_0 \right)\right]^{|m|}.
\]
Here the coefficients $\gamma^r=-1$ and $\gamma^t=\cos\theta/\cos\theta^{\prime}$ account
for the deformations of the vortex in the secondary beams: charge flip in the reflected
beam and an elliptic deformation of the transmitted beam. As a result, the centers of
gravity of the reflected and transmitted vortex beams experience the IF shift
\begin{equation}
\langle Y^{r} \rangle_{2}=m\frac{D}{k}\langle K_{x}^{r} \rangle_0,~
\langle Y^{t} \rangle_{2}=-m\frac{\cos\theta^{\prime}}{\cos\theta}\frac{D}{k}\langle K_{x}^{t} \rangle_0.
\end{equation}

The net IF shift of an arbitrarily polarized vortex beam is the sum of the contributions
(6), (10), and (11):
\begin{equation}
\langle Y^{r,t} \rangle = \langle Y^{r,t} \rangle _{0} + \langle Y^{r,t} \rangle _{1} + \langle Y^{r,t} \rangle _{2}.
\end{equation}
This is the first main result that describes the total polarization- and vortex-
dependent IF shift.

Similarly to the vortex-induced IF effect (11) associated with the angular GH shift (7),
there exists a reciprocal effect of the \textit{vortex-induced GH shift} caused by the
angular IF shift (8). In a manner, similar to the discussion above, we obtain
\begin{equation}
\langle X^{r} \rangle=-m\frac{D}{k}\langle K_{y}^{r} \rangle_0,~\langle X^{t} \rangle=m\frac{\cos\theta^{\prime}}{\cos\theta}\frac{D}{k}\langle K_{y}^{t} \rangle_0.
\end{equation}
This is the second main result of this Letter. It predicts a completely new type of the
GH shift, which occurs in the regime of partial reflection, in a sharp contrast to the
usual GH effect (4). The vortex-induced GH effect vanishes for $p$, $s$, and circular
polarizations of the incident beam and reaches maximal values for linear polarizations
inclined at $\pi/4$ angles: $\chi=\pm 1$.

It can be shown that the \textit{angular} GH and IF shifts, Eqs. (7) and (8), acquire
additional factor $(1+|m|)$ in the case of incident vortex beam:
\begin{equation}
\langle K_{x,y}^{r,t} \rangle = (1+|m|)\langle K_{x,y}^{r,t} \rangle_0.
\end{equation}

The above equations (4)--(8) and (10)--(14) describe \textit{all} the main shifts of
polarized vortex beams with an axially-symmetric intensity profile. One may observe that
the angular shifts (7), (8), and (14) fulfil conservation laws for the $x$ and $y$
components of the total linear momentum in the system~\cite{Fed2}:
\begin{eqnarray}
\nonumber
- Q^r \langle K_{x}^{r} \rangle \cos\theta  + Q^t \langle K_{x}^{t} \rangle \cos\theta^{\prime} = 0,\\
Q^r \langle K_y^{r} \rangle + Q^t \langle K_y^{t} \rangle = 0,
\end{eqnarray}
whereas linear IF shifts (6), (10)--(12) fulfil conservation law for the $z$ component of
the total AM \cite{Player,Fed1988,Bliokh1,Bliokh2,Fed1}:
\begin{eqnarray}
Q^r (J_z^r-\langle Y^{r} \rangle k\sin\theta) + Q^t (J_z^t-\langle Y^{t} \rangle k\sin\theta) =J^i_z.
\end{eqnarray}
Here $J^{a}_z$ is the $z$-components of the total intrinsic AM of the $a$-th beam, which
consist of orbital (vortex) and spin (polarization) contributions:
\begin{eqnarray}
{\bf J}^{a} = {\bf L}^{a} + {\bf S}^{a},~{\bf S}^{a} = \sigma^a {\bf \hat Z}^a,
\end{eqnarray}
where $\sigma^i = \sigma$ and $\sigma^{r,t}=2 {\rm Im}({e_{\parallel}^{r,t}}^*
e_{\perp}^{r,t})$ are the helicities of the corresponding beams.

\begin{figure}[t]
\centering \scalebox{0.35}{\includegraphics{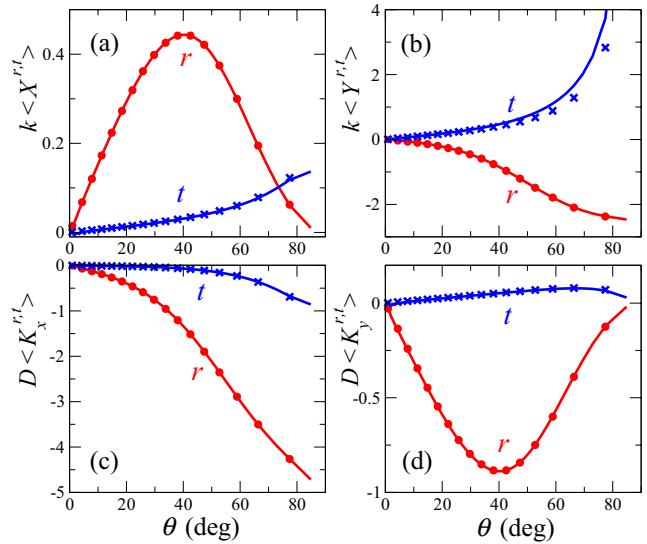}} \caption{(Color online)
Numerical (curves) and theoretical (symbols) results for the linear and angular IF and GH
shifts of the reflected ($r$) and transmitted ($t$) vortex beam with complex
polarization, as a function of the angle of incidence $\theta$. Parameters are: $n=1.5$,
$\mu=1$, $e_{\parallel}=1/\sqrt{3}$, $e_{\perp}=(1+i)/\sqrt{3}$, and $m=1$. For the
wavelength of 632nm, $\langle X^r \rangle=44$nm for $\theta=40$deg and $\langle Y^r
\rangle=-240$nm for $\theta\simeq 90$deg. These values are proportionally enhanced for
higher $m$.} \label{Fig2}
\end{figure}

To verify our results, we perform numerical simulations of the problem based on the
plain-wave decomposition of the incident polarized vortex beam and Fresnel-Snell's
formulas applied to each plane-wave component of the beam spectrum. The numerically
calculated coordinates of the centers of gravity of the reflected and transmitted beams
in the wave-vector and coordinate space are compared with analytical expressions
(6)--(8), (10)--(14) in Fig. 2, which shows an excellent agreement.

In conclusion, we have calculated linear and angular GH and IF shifts arising upon the
reflection and transmission of a polarized vortex beam at an interface between two media.
We have established a close relation between the spin-induced angular shifts and
orthogonal vortex-induced linear shifts, and have predicted a new type of the GH shift
for a vortex beam, which occurs in the partial-reflection regime when the GH shift for
Gaussian beams vanishes. The value and sign of the vortex-induced shifts can be
controlled by the vortex charge.


\end{document}